\documentclass{spring}

\usepackage[english]{babel}
\usepackage{booktabs}
\usepackage{multirow}
\usepackage{graphicx}
\usepackage{siunitx}

\springyear{2026}
\springdate{April 21--22}
\springlocation{Heidelberg, Germany}

\usepackage{tcolorbox}

\tcbuselibrary{skins}
\newtcolorbox{takeaway}{
  enhanced,
  arc=0pt,
  outer arc = 0pt,
  sharp corners,
  boxsep=0pt,
  colback=black!15!white, 
  colframe=black!30!white, 
  leftrule = 5mm,
  toprule = 0mm,
  bottomrule = 0mm,
  rightrule = 0mm,
  overlay={%
    \node[black, anchor=north, rotate=90] at (frame.west) {{\textbf{Take Away}}};
  }
}

\usepackage{soul}
\DeclareRobustCommand{\tightcolorbox}[2]{{\sethlcolor{#1}\hl{#2}}}

\definecolor{alertred}{rgb}{0.65,0.02,0.01}

\usepackage{calc}

\title{On the Challenges of Holistic Intrusion Detection in ICS}

\author{
  Stefan Lenz\textsuperscript{1},
  Julia Raab\textsuperscript{1}, 
  Benedikt Holzbach\textsuperscript{1}, 
  Deniz Köller\textsuperscript{1},
  Sotiris Michaelides\textsuperscript{1},
  Martin Henze\textsuperscript{1,2} \\
  \textsuperscript{1}Security and Privacy in Industrial Cooperation, RWTH Aachen University \;\textbullet{}\;
  \textsuperscript{2}Cyber Analysis\\ \& Defense, Fraunhofer FKIE  \;\textbullet{}\;
 \email{\{lenz, michaelides, henze\}@spice.rwth-aachen.de}, \\ \email{\{julia.raab, benedikt.holzbach, deniz.koeller\}@rwth-aachen.de} 
}

\begin{document}

\maketitle
\thispagestyle{fancy}

\begin{abstract}
  Past attacks against industrial control systems (ICS) show that adversaries often target both the ICS network \emph{and} the physical process to achieve potential catastrophic impact.
  To secure ICS, intrusion detection systems promise timely uncovering of such adversaries.
  However, as these detection mechanisms typically focus on isolated characteristics of ICS (e.g., packet timings), multiple detection systems have to be deployed in parallel, complicating their operation in practice. 
  In this work, to spur discussion and further research, we present challenges encountered during our research towards a holistic intrusion detection system aiming to cover all dimensions of an ICS.
\end{abstract} 

\section{Holistic\,Intrusion\,Detection\,in\,ICS}

\begin{figure}
\includegraphics{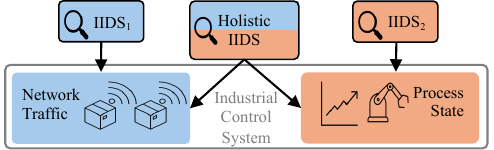}
\caption{Holistic industrial intrusion detection systems (IIDS) aim to monitor both network behavior and physical process state, while traditional IIDS focus on one.}
\label{fig:holistic}
\vspace{-1.5em}
\end{figure}

By connecting industrial control systems (ICS) to the Internet, these safety critical systems are exposed to sophisticated attackers~\cite{knapp2024icsHistory, attacks_report, bader2023wattson}.
As ICS consist of low-resource legacy devices that are difficult to replace and not capable of complex security operations, retro-fitting attack detection into ICS has been a major focus of recent research~\cite{wolsing2022ipal, lamberts2023evaluations}.
Since the behavior of ICS is tightly coupled with a deterministic physical process, typical industrial intrusion detection systems (IIDS) model benign behavior (e.g., packet timings~\cite{lin2018IaT} or process states~\cite{wolsing2022simple}) to later identify anomalies. 
However, as attacks materialize in the network \emph{and} the physical dimension of ICS, comprehensive detection must cover all aspects of ICS~\cite{wolsing2023ensemble}.
Thus, we argue that IIDS should aim for a holistic detection approach, covering multiple dimensions simultaneously as depicted in Fig.~\ref{fig:holistic}.

One method to achieve holistic detection is to use multiple one-dimensional detectors in parallel~\cite{wolsing2023ensemble}.
Although such ensembles promise great detection performance \emph{in theory}, they are difficult to deploy, since each detector requires sufficient training data as well parameter optimization.
Furthermore, individual detectors risk to miss interplay between different dimensions and it is unclear how to combine alerts of individual detectors in a sensible manner~\cite{wolsing2023ensemble}.

To overcome these challenges, we set out to address the need for a \textit{practical} holistic detection method. 
Instead of combining different detectors specializing in individual aspects, we strive to cover the complete benign behavior of an ICS in one holistic model.
During this pursuit, we encountered several challenges while exploring different approaches.
In this paper, we discuss these challenges regarding process state discretization (§\ref{sec:discretization}), IIDS parameterization (§\ref{sec:parameterization}), and collecting sufficiently good training data (§\ref{sec:training}).

\begin{figure*}[t]
   \begin{minipage}[t]{.325\textwidth}
    \vspace*{-1.5cm}

  \scriptsize
  \setlength{\tabcolsep}{0.5pt} %
  \begin{tabular}{@{} l c c c c c c c c c@{}}
        \toprule
        & \multicolumn{9}{c}{\textbf{Metric}} \\ \cmidrule(lr){2-10}
        \textbf{Name} & \rotatebox{90}{Accuracy} & \rotatebox{90}{Precision} & \rotatebox{90}{Recall} & \rotatebox{90}{F0.1} & \rotatebox{90}{F1} & \rotatebox{90}{eTaP} & \rotatebox{90}{eTaR} & \rotatebox{90}{eTaF0.1} & \rotatebox{90}{eTaF1}\\
        \midrule
        Jenkspy &  0.68 & 0.23 &  0.15 &  \cellcolor{green!50} 0.23 &  \cellcolor{green!50} 0.21 & 0.23 & \cellcolor{green!50} 0.52 & \cellcolor{green!50}0.24 & \cellcolor{green!50} 0.32 \\
        kMeans &  0.68 &  0.24 & \cellcolor{green!50} 0.16 &\cellcolor{green!50} 0.23 & \cellcolor{green!50} 0.21 & 0.24 & \cellcolor{green!50} 0.52 & \cellcolor{green!50} 0.24 & \cellcolor{green!50} 0.32 \\
        Quantiles & \cellcolor{green!50} 0.77 & \cellcolor{green!50} 0.71 & 0.00 & 0.03 & 0.00 & \cellcolor{green!50} 0.69 & 0.14 & 0.66 & 0.23\\
        \bottomrule
  \end{tabular}%
   \end{minipage}%
   \hfill%
    \begin{minipage}[t]{.325\textwidth}
      \centering
      \vspace*{-1.5cm}
      \includegraphics[width=\linewidth]{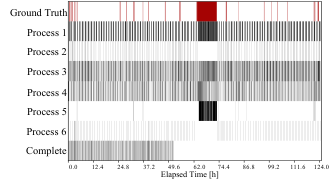}
   \end{minipage}%
   \hfill%
   \begin{minipage}[t]{.325\textwidth}
      \centering
      \vspace*{-1.5cm}
      \scriptsize
      \setlength{\tabcolsep}{3pt} %
      \begin{tabular}{l c c c }
        \toprule
         & \multicolumn{3}{c}{\textbf{Inter-Arrival Times}}\\ 
        \cmidrule(lr){2-4}
        \textbf{Medium} & L1 [ms] & L2 [ms] & L3 [ms] \\
        \midrule
        Wired & 46.06 (0.37) & 46.06 (0.37) & 100.00 (0.02) \\
        Wireless$^{+}$ & 46.06 (0.37) & 46.06 (0.37) & 100 (0.12)\\
        Wireless$^{-}$ & 45.97 (1.47) &  45.97 (1.56) & 100 (0.67)\\
        \bottomrule
      \end{tabular}
   \end{minipage}%

   \vspace{0.5em}
   \begin{minipage}[t]{.32\textwidth}
      \captionof{table}{The best algorithm (\tightcolorbox{green!50}{green}) to discretize process data not only depends on data distribution but also on the detection performance metric. 
      }
      \label{tb:discretization}
  \end{minipage}%
  \hfill%
  \begin{minipage}[t]{.32\textwidth}
      \captionof{figure}{Alerts ({\color{white}\tightcolorbox{black}{black}}) of LLM detection on SWaT ({\color{white}\tightcolorbox{alertred}{red}}) are sensitive to process data variance. Monitoring subprocesses reduces processing overhead.}      
      \label{fig:parameterization}
  \end{minipage}%
  \hfill%
  \begin{minipage}[t]{.32\textwidth}
      \captionof{table}{Differences in inter-arrival times in wired and good ($^{+}$) / bad ($^{-}$) wireless channels show the need for detection to adapt to dynamic behavior.}      
      \label{tb:iat}
  \end{minipage}%
  \vspace{-1em}
\end{figure*}

\section{Challenge 1: Discretization}
\label{sec:discretization}

To create a holistic IIDS capturing the complete ICS in a single model, we explored the use of \emph{process mining}, which has shown promising results in modeling physical process state~\cite{myers2018ids,apolinario2024fingerCi} as well as network traffic~\cite{wakup2015pcapMining}.
Indeed, initial experiments showed that such a detector can detect attacks targeting both these dimensions, while performing \textit{on par} with other IIDSs that capture only one dimension.
However, as process mining relies on discrete labels (e.g., bins) to incorporate the physical state into the model, the fundamental \emph{challenge of discretization} arises to assign such labels to continuous ICS data (e.g., temperatures or fill levels).

Although many discretization mechanisms, such as statistical methods or unsupervised learning approaches (e.g., clustering), solve this problem, the quality of the resulting labels/clusters is highly dependent on the distribution of the underlying data~\cite{vanderAalst2020processMiningChallenge,fraley1998Clustering,wegman2021clustering}.
To choose the optimal discretization method for an IIDS using process mining, we compared multiple discretization methods based on their resulting \textit{detection performance} as summarized in Tb.~\ref{tb:discretization}. 
Our experiments show that, even while utilizing benchmarking datasets (with posterior knowledge that is not achievable in any practical scenario), this task is not trivial, as the best discretization method (\tightcolorbox{green!50}{green} in Tb.~\ref{tb:discretization}) depends on the considered performance metric.

\begin{takeaway}
  While discretization mechanisms promise well-defined labels for process data, in the context of intrusion detection, identifying the ``best'' approach depends on the considered detection metric. 
\end{takeaway}

\section{Challenge 2: Parameterization}
\label{sec:parameterization}

A holistic IIDS, e.g., based on process mining, requires multiple additional parameters (e.g., length of the ICS cycle in packets) that must be set manually, resulting in a need for hard-to-come-by expert knowledge.
To address the resulting \emph{challenge of parameterization}, we explored different ``closed-box'' modeling approaches (e.g., based on LLMs), which promise no parameterization at all. 
The goal of such approaches is to gain ``adequate'' detection performance while utilizing the reasoning capabilities of LLMs. %
Exemplary experiments on the SWaT~\cite{swatDataset} dataset (cf.\ Fig.~\ref{fig:parameterization}) indicate that an LLM-based IIDS struggles to monitor the complete process in a holistic manner, due to excessive resource demands, even exceeded the capabilities of the high-performance cluster node (96GB VRAM) we utilized.

Therefore, we focused on individual subprocesses in further experiments.
While the smaller scope reduces computational complexity, the detector provided nearly-constant alerts, rendering attacks unrecognizable (except for Process 5 which has comparatively low variance).
Still, due to its closed-box nature, it is impossible to comprehend the detector's reasoning behind alerts~\cite{skrodelis2026HybdridIcs}, making discerning between true and false alarm challenging.
Together with the high resource demands, these issue prevent a practical application of a parameterless approach as a holistic IIDS.

\begin{takeaway}
  Although addressing the challenge of parameterization, monolithic closed-box detectors face challenges in deployment, resource usage, and interpretability, preventing their use for holistic IIDS.
\end{takeaway}

\section{Challenge 3: Good Training Data}
\label{sec:training}

To meet increasing demands for flexibility and efficiency, ICS adopt new technologies such as wireless communication~\cite{michaelides2025industry5G}.
These more dynamic systems exaggerate the \emph{challenge of gathering good training data} in IIDS research~\cite{wolsing2024deployment}.

To investigate this issue, we built an ICS simulation enabling us to change the communication medium without altering the process behavior~\cite{lenz2026swics}.
There, we analyze a popular timing-based IIDS~\cite{lin2018IaT}, which uses deterministic inter-arrival timings (i.e., the time between two packets) to discern between benign and anomalous behavior for three communication scenarios: \textit{Wired} (as a baseline), \textit{Wireless$^{+}$} (a wireless channel in perfect condition), and  \textit{Wireless$^{-}$} (a wireless channel experiencing substantial disturbance).

Tb.~\ref{tb:iat} shows the mean ($\mu$) inter-arrival times and standard deviation ($\sigma$) for three links from our simulated ICS.
First, these results show that a wireless channel in perfect conditions can achieve the same link quality as the wired medium, although the variance in some links may be higher (Tb.~\ref{tb:iat} L3).
The wireless medium under disturbance, however, cannot provide such link quality.
Although the mean inter-arrival time is similar or even lower than in the other scenarios, $\sigma$ is substantially higher for all links.
This behavioral change can trigger false alerts in timing-based IIDS, which cannot distinguish a distressed channel from an attack.

As the quality of a wireless channel can change anytime (e.g, due to noise), the model of the IIDS might not reflect the actual (communication) behavior of the ICS, thus either resulting in false alarms or missed attacks.
Additionally, current research~\cite{schuster2025IcsTraffic} also suggests that ICS behavior might not be as deterministic as often assumed.
Thus, IIDS research should also focus on mechanisms that can cope with these dynamic environments, especially for holistic IIDS. 

\begin{takeaway}
  Even though the collection of suitable training data is a known challenge for intrusion detection, modern, e.g., wireless, ICS with more dynamic behavior further complicate this challenge.
\end{takeaway}

\section{Conclusion}

To secure ICS against multi-faceted threats targeting both physical process \emph{and} communication, we pursue a holistic detection approach which covers both dimensions of an ICS simultaneously.
In this paper, we report on our research process and discuss challenges---regarding process state discretization (§\ref{sec:discretization}), parameterization (§\ref{sec:parameterization}), and training data (§\ref{sec:training})-- we faced to push such a holistic IIDS towards practical viability.
Ultimately, our findings identify ample research opportunities, especially highlighting opportunities in more ``dynamic'' detection mechanisms capable of handling characteristics of future ICS.

\balance{}

\paragraph{Acknowledgements}

The research underlying this publication has in parts been funded by the German Federal Ministry of Research, Technology and Space (BMFTR) under funding reference number 16KIS2409K (6GEM+).
Computations were performed with computing resources granted by RWTH Aachen University under project thes2001.
The authors are responsible for the content of this publication.

\printbibliography{}

\end{document}